\documentclass[letter,longauth]{aa}
\usepackage{graphicx,color}
\usepackage{amssymb,verbatim,hyperref}
\usepackage{txfonts}
\usepackage{natbib}
\usepackage{mathptmx}

\def\lxm{$L_X-M$}
\def\lxt{$L_X-T$}
\def\fgasm8{fgas$-8\sigma$}

\begin{document}

\title{The impact of strong feedback on galaxy group scaling relations}
\author{D. Eckert\inst{\ref{gva}} \and R. Seppi\inst{\ref{gva}} \and J. Braspenning\inst{\ref{mpia}} \and A. Finoguenov\inst{\ref{uhel}} \and F. Gastaldello\inst{\ref{iasfmi}} \and L. Lovisari\inst{\ref{iasfmi},\ref{cfa}} \and E. O'Sullivan\inst{\ref{cfa}} \and S. Ettori\inst{\ref{oabo},\ref{infn}} \and B. D. Oppenheimer\inst{\ref{ucol}} \and M. A. Bourne\inst{\ref{hertf},\ref{kavli}} \and D.-W. Kim\inst{\ref{cfa}} \and M. Sun\inst{\ref{ualab}} \and H. Khalil\inst{\ref{uhel}} \and G. Gozaliasl\inst{\ref{aalto}} \and Y. E. Bahar\inst{\ref{oabo}} \and V. Ghirardini\inst{\ref{oabo}} \and W. Cui\inst{\ref{uautmad},\ref{ciaff},\ref{ued}} \and K. Kolokythas\inst{\ref{rhodes},\ref{sarao}} \and S. McGee\inst{\ref{ubirm}}}
\institute{
Department of Astronomy, University of Geneva, Ch. d'Ecogia 16, 1290 Versoix, Switzerland\label{gva}\\
\email{Dominique.Eckert@unige.ch}
\and
Max-Planck-Institut für Astronomie, Königstuhl 17, D-69117 Heidelberg, Germany\label{mpia}
\and
Department of Physics, University of Helsinki, Gustaf Hällströmin katu 2, 00560 Helsinki, Finland\label{uhel}
\and
INAF - IASF Milano, Via Alfonso Corti 12, 20133 Milan, Italy\label{iasfmi}
\and
Center for Astrophysics | Harvard \& Smithsonian, 60 Garden Street, Cambridge, MA 02138, USA\label{cfa}
\and 
INAF – Osservatorio di Astrofisica e Scienza dello Spazio di Bologna, Via P. Gobetti 93/3, 40129 Bologna, Italy\label{oabo}
\and
INFN, Sezione di Bologna, viale Berti Pichat 6/2, 40127 Bologna, Italy \label{infn}
\and
CASA, Department of Astrophysical and Planetary Sciences, University of Colorado, 389 UCB, Boulder, CO 80309, USA\label{ucol}
\and
Centre for Astrophysics Research, Department of Physics, Astronomy and Mathematics, University of Hertfordshire, College Lane, Hatfield, AL10 9AB, UK\label{hertf}
\and
Kavli Institute for Cosmology, University of Cambridge, Madingley Road, Cambridge CB3 0HA, UK\label{kavli}
\and
Department of Physics and Astronomy, University of Alabama in Huntsville, Huntsville, AL35899, USA\label{ualab}
\and
Department of Computer Science, Aalto University, PO Box 15400, Espoo, FI-00 076, Finland\label{aalto}
\and
Departamento de F\'isica Te\'orica, M-8, Universidad Aut\'onoma de Madrid, Cantoblanco, E-28049, Madrid, Spain. \label{uautmad}
\and
Centro de Investigaci\'on Avanzada en F\'isica Fundamental (CIAFF), Universidad Aut\'onoma de Madrid, Cantoblanco, E-28049 Madrid, Spain. \label{ciaff}
\and
Institute for Astronomy, University of Edinburgh, Royal Observatory, Edinburgh EH9 3HJ, United Kingdom \label{ued}
\and
Centre for Radio Astronomy Techniques and Technologies, Department of Physics and Electronics, Rhodes University,\\ P.O. Box 94, Makhanda 6140, South Africa\label{rhodes}
\and
South African Radio Astronomy Observatory, Black River Park North, 2 Fir St, Cape Town, 7925, South Africa\label{sarao}
\and 
School of Physics and Astronomy, University of Birmingham, Birmingham, B152TT, UK\label{ubirm}
}

\abstract{
Feedback from active supermassive black holes alters the distribution of matter in the Universe by injecting energy in the neighbouring hot gaseous medium, which leads to ejection of gas from the halos of galaxy groups and massive galaxies. Recent cosmological simulations such as FLAMINGO calibrate their feedback model on the baryon fractions of galaxy groups to tune the efficiency of gas ejection. However, recent observational constraints from optically selected groups and the kinetic Sunyaev-Zel'dovich effect yield lower baryon fractions than previous studies, which indicates that feedback may be more ejective than previously thought. In this work, we confirm that the scaling relations of local galaxy groups in the mass range $10^{13}-10^{14}M_\odot$ favour the fiducial FLAMINGO feedback calibration. We study the X-ray luminosity-temperature relation in a sample of 44 galaxy groups with high-quality \emph{XMM-Newton} observations. We show that highly ejective models under-predict the luminosity of galaxy groups at fixed mass at high significance ($5.7\sigma$). This discrepancy cannot be explained by selection effects and is obtained from directly measurable and minimally correlated quantities. We point out that turning observable quantities into gas fraction estimates is challenging, especially in the context of stacking large samples of heterogeneous systems. We argue that validating feedback models against observable scaling relations is necessary to warrant the validity of feedback implementations in cosmological simulations.}

\maketitle
 
\section{Introduction}

Feedback from active galactic nuclei (AGN) is a  necessary ingredient in galaxy evolution models, as it regulates the star formation in massive galaxies and reproduces the shape of the galaxy stellar mass function \citep[e.g.][]{Silk:1998}. However, implementation of this feedback differs substantially from one simulation to another, and various combinations of sub-grid parameters can yield similar stellar mass functions \citep[e.g.][]{Schaye:2023}. In contrast, the feedback models make distinct predictions for the heating of the hot atmospheres of massive halos \citep{McCarthy:2010}. Galaxy groups in the mass range $M_{500}\sim10^{13}-10^{14}M_\odot$ are particularly sensitive to feedback, as the total AGN energy input rivals or even exceeds the gas binding energy \citep{Eckert:2021}, which leads to gas ejection from the halo and sub-cosmic baryon fractions. As such, the efficiency of gas ejection can be tuned by calibrating the feedback models to match the fraction of baryons remaining within the halos \citep{McCarthy:2017,Henden2018}. In turn, these simulations can be used to predict the impact of baryonic feedback on the matter distribution on cosmological scales, which represents a leading source of systematic uncertainty for upcoming cosmological surveys \citep{Chisari:2019}.

While tuning the feedback model to reproduce the gas fractions of galaxy groups provides an effective way of implementing gas ejection within simulations, the resulting predictions require the calibration datasets to be highly accurate. In particular, the fiducial FLAMINGO model for baryonic feedback was calibrated on an heterogeneous collection of gas fraction measurements \citep[e.g.][]{Sun:2009a,Lovisari:2015,Akino:2022}. In many cases, the halo masses were estimated while assuming hydrostatic equilibrium (HSE), which is likely biased in dynamically active systems \citep[e.g.][]{Gianfagna:2021}. Moreover, X-ray surveys preferentially select the brightest, gas-rich systems. For this reason, the FLAMINGO suite includes multiple runs from the same initial condition with different feedback strengths, tuned to reproduce the mean $f_{\rm gas}-M_{500}$ relation of clusters with $M_{500}\gtrsim10^{14}M_\odot$ as well as variants offset by +$2\sigma$ (weaker feedback), -$2\sigma$, -$4\sigma$, and -$8\sigma$ (stronger feedback). 

The general understanding of gas ejection in massive halos has been rapidly evolving recently. For instance, the \emph{eROSITA} all-sky survey has detected hundreds of galaxy groups \citep{Bahar:2024,Siegel:2025}. Stacking of optically selected systems also provides an alternative view of the gas content of groups that is independent of X-ray selection \citep{Comparat:2022,Popesso:2024a,Popesso:2024b,Zhang:2024}. Independent constraints have also been obtained from the kinetic Sunyaev–Zel’dovich (kSZ) effect for halos of lower masses and higher redshifts \citep{Schaan:2021,Guachalla:2025,Hadzhiyska:2025,Lucie-Smith:2025}. These recent results seem to imply efficient gas ejection from group-scale halos and lower gas fractions than previously thought in the group mass range. When combining these measurements with galaxy-galaxy lensing estimates, it was shown that the latest measurements are consistent with the FLAMINGO models with the strongest feedback \citep[\fgasm8,][]{McCarthy:2025,Siegel:2025}. If confirmed, these results would imply a marked suppression of the matter power spectrum on $\lesssim 1$ Mpc scales \citep{Bigwood:2024}. 

In this work, we study the X-ray luminosity-temperature relation in a sample of 44 nearby galaxy groups from the \emph{XMM-Newton} Group AGN Project \citep[X-GAP,][]{XGAP}. We then compare the retrieved $L_X-T$ relation with the predictions of FLAMINGO simulations with varying feedback. Throughout the paper, we assume a concordance cosmology with $H_0=67.8$ km/s/Mpc and $\Omega_m=0.3075$ \citep{Planck:2016}.

\section{The X-GAP sample}
\label{sec:sample}

X-GAP \citep{XGAP}\footnote{\href{https://www.astro.unige.ch/xgap}{https://www.astro.unige.ch/xgap}} is a sample of 49 galaxy groups selected from the All-sky X-ray Extragalactic Sources sample \citep[AXES-SDSS;][]{Damsted:2024}. AXES was selected by cross-correlating spectroscopic galaxy groups from the Sloan Digital Sky Survey (SDSS) with at least five member galaxies \citep{Tempel:2017} with weak extended X-ray sources from the \emph{ROSAT} all-sky survey. \citet[hereafter S25]{Seppi:2025} studied the selection function of the sample through detailed semi-analytic simulations and showed that the survey selection closely resembles an X-ray flux selection with a flux limit of $\sim10^{-12}$ erg/cm$^2$/s in the 0.1-2.4 keV band. Unlike usual X-ray samples, the cross-correlation with optical groups does not preferentially select centrally peaked systems. X-GAP was then selected as a complete sub-sample of AXES with the following criteria: \emph{i)} $z\leq0.05$; \emph{ii)} number of galaxies greater than eight; \emph{iii)} $R_{500}$ smaller than $15$ arcmin. The third criterion can be written as an upper bound on the group luminosity, which reads
\begin{equation}
    \log L_{0.5-2, 500}(z) \leq 43.65 + 4.69 \log\left(\frac{z}{0.05}\right).
    \label{eq:lxsel}
\end{equation}
These criteria yielded a sample of 45 groups in the mass range $10^{13}-10^{14}M_\odot$, all of which were followed up with \emph{XMM-Newton} to obtain at least 20,000 net source counts per group (proposal ID 090389). The final sample includes four additional systems in the redshift range $0.05-0.06$ with similar observational depth. These four systems do not formally fulfil the selection criteria, and thus we do not consider them here. Finally, for one system, SDSSTG 16386, the association with the X-ray source was found to be incorrect, and the optical group is likely spurious. The detected \emph{ROSAT} source matches another group at a slightly higher redshift (SDSSTG 24595 at $z=0.058$). Finding one such source is not surprising, as the purity of the optical group finder is about 65\% (see S25). Our final selection contains 44 systems, which represents a complete sub-sample of AXES given the above criteria. 

The \emph{XMM-Newton} data were homogeneously analysed using the procedure described in \citet{Eckert:2025}, which is outlined in Appendix \ref{sec:red} and \ref{sec:integ}. The median temperature of the sample is 1.13 keV, with 16$^{th}$ and 84$^{th}$ percentiles of 0.86 and 1.82 keV. 

\begin{figure*}
    \resizebox{\hsize}{!}{\includegraphics[width=0.5\textwidth]{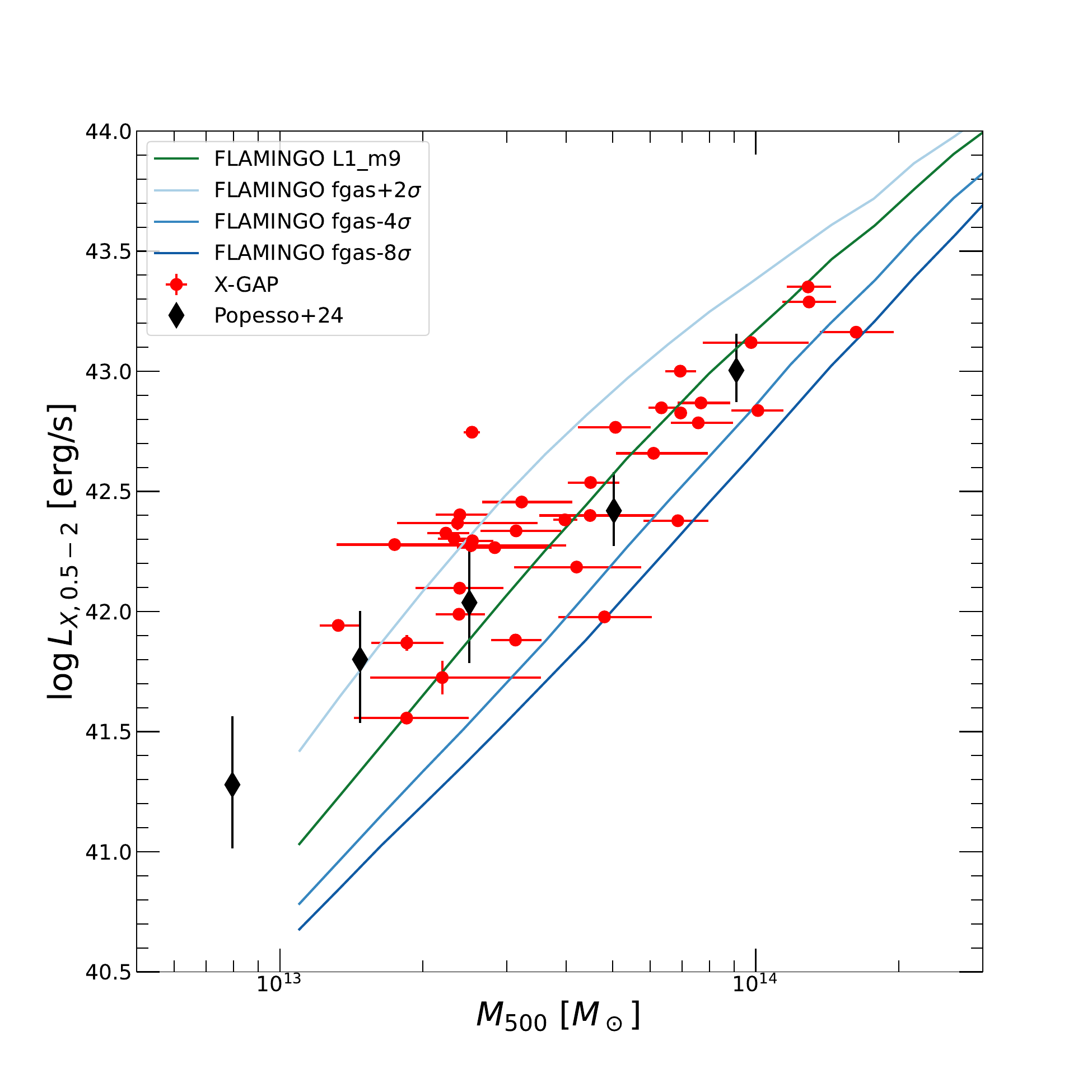}
    \includegraphics[width=0.5\textwidth]{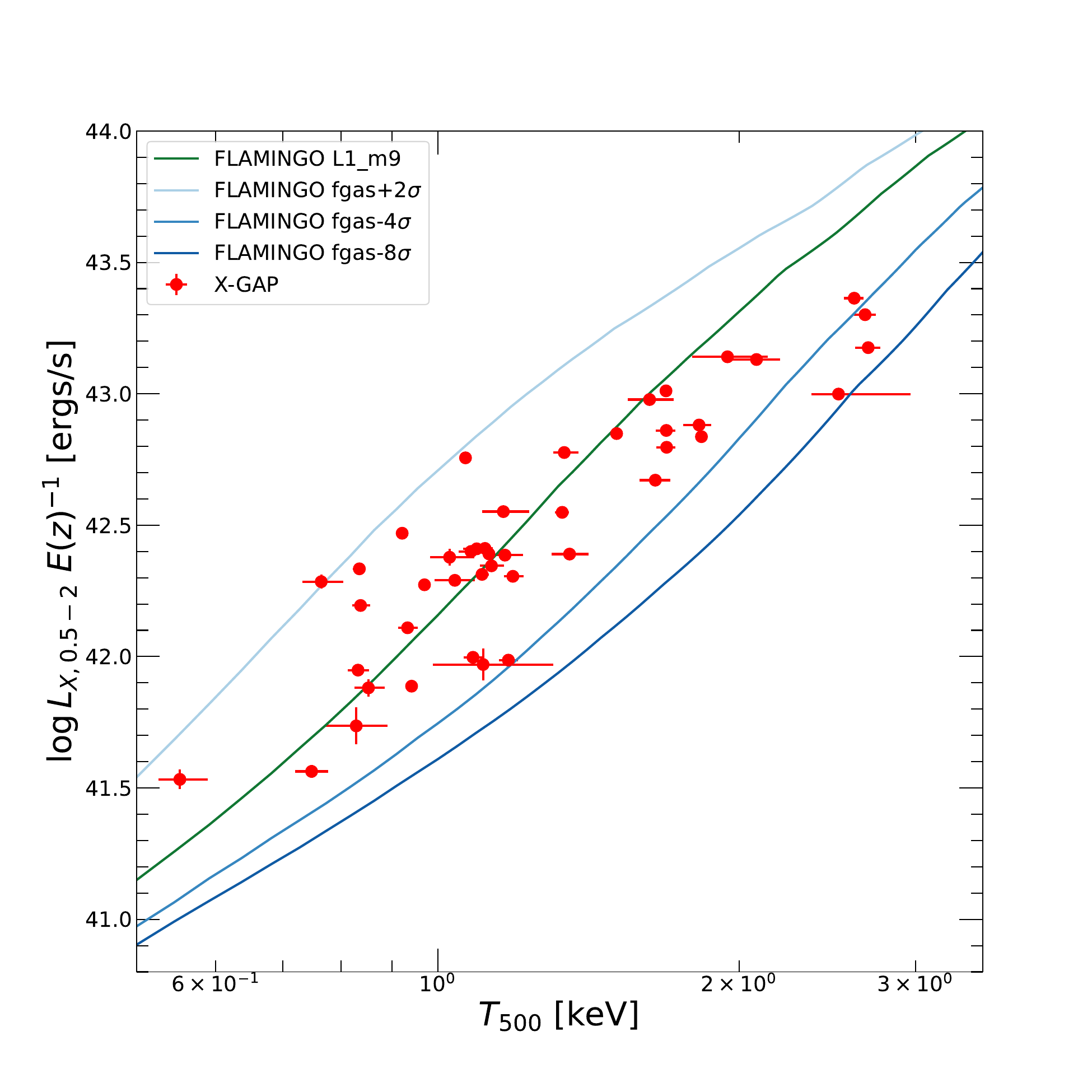}}
    \caption{Luminosity-mass (left) and luminosity-temperature (right) relations for X-GAP groups (colored symbols, see Table \ref{tab:lxt}). The solid curves show the relations obtained in various FLAMINGO runs (see B24). In the left-hand panel, the black diamonds show the luminosity-mass relation of optically selected groups in the eFEDS field \citep{Popesso:2024b}.}
    \label{fig:main}
\end{figure*}

\section{Results}
\label{sec:results}
\subsection{Luminosity-temperature and luminosity-mass relations}

In Fig. \ref{fig:main} we plot the luminosity-mass (\lxm) and luminosity-temperature (\lxt) relations for the 44 groups that strictly match the selection criteria. For the \lxm~relation, we only consider the systems with direct HSE masses (see Sect. \ref{sec:integ}), whereas the \lxt~plot includes all systems. 

We compared our results with the \lxm~relation obtained by stacking optically selected groups in the eFEDS field \citep{Popesso:2024b}. Importantly, our \lxm~relation agrees with the relation retrieved from \emph{eROSITA} stacks. The optical selection is expected to be more complete, albeit less pure, than our optical/X-ray matching technique (S25). This demonstrates that, on average, X-GAP groups are not over-luminous for their mass. For comparison, we plotted the \lxm~relations obtained for FLAMINGO simulations with varying feedback \citep[hereafter B24]{Braspenning:2024}. We considered only the nominal thermal feedback model whereby the energy injected by AGN is distributed isotropically by raising the temperature of neighbouring particles \citep{Booth:2009}. A full comparison between the properties of X-GAP and the FLAMINGO predictions is deferred to a subsequent paper (Seppi et al. in prep.). 

One can see that the \lxm~relation of the X-GAP groups agrees with the fiducial FLAMINGO run (L1\_m9), whereas all the groups lie above the FLAMINGO \fgasm8~curve. However, the \lxm~relation is highly sensitive to the accuracy of  the halo mass estimates such that an under-estimation of the halo masses may yield a better agreement with the strong feedback models. For this reason, we focused our analysis on the \lxt~relation, which is essentially insensitive to mass estimates. The X-GAP \lxt~relation agrees with the FLAMINGO fiducial run and is strongly discrepant with the models with the strongest feedback. At 1 keV ($\sim3\times10^{13}M_\odot$), the median luminosity of the X-GAP groups exceeds the prediction of the \fgasm8~run by about 0.5 dex. This indicates that this run overly reduces the gas content in the inner regions ($\lesssim0.5R_{500}$), leading to lower X-ray luminosities at fixed temperature. None of the groups fall below the median of the \fgasm8~run, although a few objects at the high-temperature end seem to favour a slightly stronger feedback than the fiducial run. On average, only $\sim15\%$ of the measured luminosity originates from the core ($<0.15R_{500}$) such that this conclusion is not driven exclusively by the behaviour of the inner regions.

\subsection{Modelling selection effects}
\label{sec:selfunc}

A key point in the comparison between the X-GAP groups and the FLAMINGO predictions is the potential impact of selection effects on our sample. In case our selection procedure preferentially includes X-ray bright systems and misses an important population of low-luminosity systems, the \lxt~and \lxm~relations of the selected systems will be biased with respect to the mean relation. 

To assess the impact of sample selection on the retrieved scaling relations, we ran Monte-Carlo simulations to study the expected properties of the selected samples in the various FLAMINGO runs. We drew objects from the halo mass function, generated observable properties given the median relations and their intrinsic scatter (B24), and applied the survey selection function calibrated in S25 from full SDSS and RASS mocks. We used the public code \texttt{colossus} \citep{Diemer:2018} to compute the halo mass function, and randomly drew halo masses within an area equivalent to SDSS (7,480 deg$^2$) and the redshift range of interest ($0.01<z<0.05$). We then assigned a luminosity and a temperature to each halo using the FLAMINGO scaling relations, assuming a log-normal intrinsic scatter. We applied an upwards correction of 0.1 dex to the luminosities to account for the contribution of projected gas along the line of sight, which corresponds to the projected contribution of a beta-model with $\beta=0.4$ \citep[typical of galaxy groups, e.g.,][]{Spinelli:2025}. Finally, we used the AXES selection function to calculate the detection probability, and populate the selected sample through rejection sampling. We then applied the X-GAP selection criteria (see Sect. \ref{sec:sample} and Eq. \ref{eq:lxsel}) to extract mock X-GAP-like samples. 

\begin{figure*}
    \resizebox{\hsize}{!}{\includegraphics{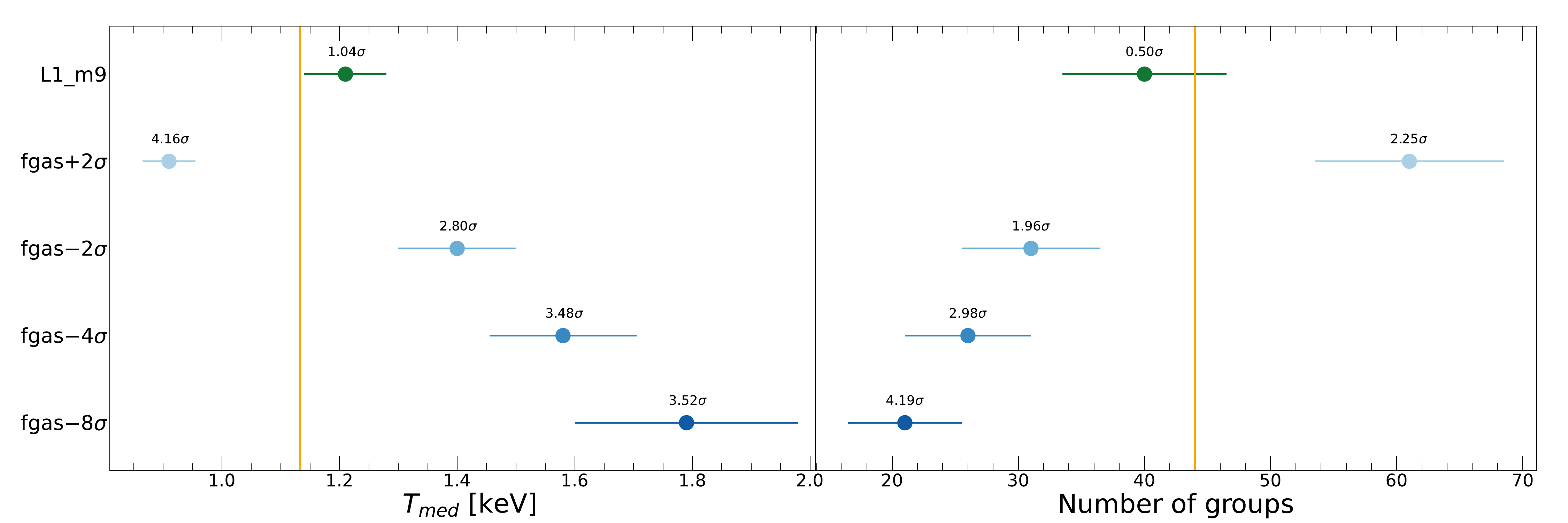}}
    \caption{Predicted median temperatures (left) and number of selected groups (right) for FLAMINGO runs with varying feedback. Each data point shows the median and 16th and 84th percentiles of simulated X-GAP-like mock samples. The orange vertical lines show the median temperature and the number of selected groups in the observed X-GAP sample. The numbers on top indicate the statistical significance of the difference with the data.}
    \label{fig:runs}
\end{figure*}

For each FLAMINGO run, we repeated the exercise 10,000 times to study the variance in the properties of the selected samples. We found that the properties of the mock X-GAP-like samples strongly depend on the feedback parameters. Most notably, the median temperature of the selected groups, $\bar T$, and the number of objects in the sample, $N_{det}$, vary strongly from one model to another. The median luminosity does not change much, as the luminosity range is set by the sample selection criteria. In Fig. \ref{fig:runs} we show the values of $\bar T$ (left) and $N_{det}$ (right) for the various FLAMINGO runs. For the fiducial run, we approximately predicted the right number of groups ($40\pm6$) and the right temperature ($1.21\pm0.07$ keV) compared to the real X-GAP sample. Runs with stronger feedback tend to under-predict the number of selected groups, as the lower luminosity at fixed mass implies that our luminosity range corresponds to more massive systems, which are rarer. For the same reason, these runs over-predict $\bar T$. In particular, the run with the strongest feedback (\fgasm8) predicts that our selection criteria should yield $21_{-4}^{+5}$ groups with $\bar T=1.79\pm0.18$ keV. We used the distributions of $\bar T$ and $N_{det}$ to quantify the discrepancy between the predictions of the FLAMINGO \fgasm8~run and the properties of X-GAP. We found that the two indicators are uncorrelated, and thus the resulting probabilities can be readily combined. Together, the two indicators are discrepant with the FLAMINGO \fgasm8~run at the $5.7\sigma$ level. While this conclusion awaits further validation from a full comparison of FLAMINGO halos with X-GAP systems (Seppi et al. in prep.), including an assessment of cosmic variance, this result shows that the X-GAP sample is weakly sensitive to selection effects.

\section{Discussion and conclusion}
\label{sec:disc}

The comparison between the X-GAP \lxm~and \lxt~relations and the FLAMINGO predictions shows that the models with the strongest feedback are in strong tension with the properties of local galaxy groups, in agreement with the original FLAMINGO calibration \citep[B24]{Kugel:2023}. This is an important conclusion, as the FLAMINGO \fgasm8~model was found to be in good agreement with the gas fractions of \emph{eROSITA} groups and the kSZ signal \citep{Siegel:2025}. In contrast, we find that the FLAMINGO models involving highly ejective feedback yield groups that are too devoid of gas in their inner regions, which leads to substantially lower X-ray luminosities at a fixed mass or temperature. This conclusion holds even when sample selection is taken into account (see Sect. \ref{sec:selfunc}). Models that are not overly ejective are required to broaden the gas distribution without overly reducing the X-ray luminosities \citep[e.g.][]{Bigwood:2025}. Alternatively, the FLAMINGO model may not exactly reproduce the true mass and redshift trends, which may still reconcile our results with the kSZ data.

We note that the luminosities predicted by simulations are affected by uncertainties in supernova yields and metal production, which impacts the estimate of the cooling function. However, B24 showed that FLAMINGO metallicities are substantially over-estimated, such that the predicted luminosities at fixed gas fraction are over-estimated. A lower and more realistic metallicity would decrease the FLAMINGO luminosities and further increase the discrepancy with the X-GAP \lxt~relation.

We point out that the inconsistency between the X-GAP \lxt~relation and the FLAMINGO \fgasm8~run involves minimal assumptions on the observables. The integrated luminosities and temperatures are both directly observable quantities, and the two measurements are minimally correlated. The two quantities depend on the estimated halo mass only through the integration aperture $R_{500}$. If the retrieved apertures are slightly underestimated, a larger integration aperture would increase the luminosities, as more flux would be included, and decrease the temperatures, as the outer regions tend to be cooler. Thus, any mass bias would move the points in the opposite direction and increase the discrepancy. The other potential sources of systematics in the measurements are very small. The spectroscopic X-ray temperatures of 1-2 keV systems should be close to the true mass-weighted value, as the bremsstrahlung cut-off falls well within the classical X-ray bandpass and the Fe-L complex acts as a temperature calibrator. Regarding the luminosity, it is the most direct observable, and the associated systematics are minimal. The uncertainties associated with absolute flux calibration, $H_0$ value, and projection effects amount at most to 0.1 dex, which is much less than the differences observed here. 

The same cannot be said of the gas fraction measurements, especially when they are derived by stacking large samples of heterogeneous systems. Substantial systematics are associated with the estimate of halo masses, which is not a direct observable of any survey. The gas mass of group-scale systems is also difficult to retrieve, as the emissivity of $\sim1$ keV plasma is highly sensitive to the metal abundance profile \citep{Lovisari:2021}. Within a single system, temperature and metallicity gradients can lead to emissivity variations by a factor of three for a fixed emission measure, which implies uncertainties of $\sim30\%$ on the retrieved gas mass. Moreover, a substantial fraction \citep[$\sim30\%$,][S25]{Robotham:2011} of optically selected groups are expected to be loose aggregations of galaxies that have not yet virialised within a common gravitational well, such that their intragroup medium is yet to be heated to X-ray emitting temperatures and the corresponding gas mass may be underestimated. Finally, the strong statistical anti-correlation between gas fraction and halo mass implies that small systematic effects propagate in a non-linear way to the estimated gas fraction at a fixed mass. 

Together, the systematic uncertainties highlighted above easily amount to a factor of two in the exact gas fraction of galaxy groups of $M_{500}\sim 3\times10^{13} M_\odot$, and therefore the gas fraction of these systems remains largely unknown. We argue that observable scaling relations provide a robust benchmark to calibrate the impact of AGN feedback on the hot atmospheres of galaxy groups, as these measurements do not rely on our knowledge of the underlying halo mass. Figure \ref{fig:main} highlights the sensitivity of the \lxt~relation to the feedback model, as small differences in gas fractions translate into strong discrepancies in the predicted luminosities at fixed temperature. Core-excised quantities can also provide more direct proxies of gas ejection in group outskirts, as they avoid the complexity of the inner regions where cooling losses are important.

\bibliographystyle{aa}
\bibliography{biblio}

\begin{acknowledgements}
Based on observations obtained with XMM-Newton, an ESA science mission with instruments and contributions directly funded by ESA Member States and NASA. DE and RS acknowledge support from the Swiss National Science Foundation (SNSF) under grant agreement 200021\_212576. LL acknowledges support from INAF grant 1.05.12.04.01. MAB is supported by a UKRI Stephen Hawking Fellowship (EP/X04257X/1).
\end{acknowledgements}

\appendix

\section{Data reduction}
\label{sec:red}

We collected the \emph{XMM-Newton} data for the 49 X-GAP groups and uniformly analysed them using the \texttt{XMMSAS} package v19.1 and the X-COP analysis pipeline \citep{Rossetti:2024}. The data analysis procedure closely follows the procedure outlined in \citet{Eckert:2025}. After performing the standard event screening procedures, we extracted photon maps from the European Photon Imaging Camera (EPIC) data in the [0.7-1.2] keV band, which maximises the signal-to-background ratio. We then computed exposure maps and non X-ray background maps following the procedure outlined in \citet{Rossetti:2024}. Finally, we extracted spectra in concentric annuli centred on the X-ray emission peak and fitted the resulting spectra in \texttt{XSPEC} with the Atomic Plasma Emission Code (APEC) v3.0.9. The \emph{XMM-Newton} non X-ray background was modelled as a combination of the quiescent particle background from filter-wheel-closed data, and residual soft protons, predicted from the difference in measured high-energy count rate between the regions located inside and outside the field of view. The local X-ray background was modelled as a combination of two APEC components describing the local hot bubble and the Galactic halo, and a hard power law to describe the cosmic X-ray background. The intensity of these components was jointly fitted to the spectra of the outermost region of the \emph{XMM-Newton} field of view and to the \emph{ROSAT} all-sky survey spectra extracted in a circular annulus located 1 and 1.5 degrees from the group centre. All the background components were then properly rescaled to the area of each annular region to determine the remaining source spectrum. For details on the data analysis procedure we refer the reader to \citet{Eckert:2025}. 

\section{Integrated quantities of X-GAP groups}
\label{sec:integ}

In most cases (41/44), the observations were deep enough to enable an analysis of the radial temperature profile of the source. In the remaining three cases, the \emph{XMM-Newton} observations were strongly affected by soft proton flares, with a clean exposure time smaller than 2 ks. 

Whenever a temperature profile is available, we fitted a Navarro-Frenk-White (NFW) model to the temperature and surface brightness profiles under the assumption of HSE, using the public packages \texttt{pyproffit} \citep{Eckert:2020} and \texttt{hydromass} \citet{Eckert:2022}. We then estimated the overdensity radius $R_{500}$ from the NFW fit. We integrated the surface brightness profile out to the derived $R_{500}$ radius to determine the total cylindrical X-ray luminosity in the rest-frame $[0.5-2]$ keV band. We also integrated the model temperature profile out to $R_{500}$ to estimate the mean spectroscopic temperature $T_{500}$. We note that the best fitting NFW masses agree well with the masses estimated from mass-temperature \citep{Umetsu:2020} and mass-velocity dispersion relations \citep{Munari:2013}, with typical differences of $\sim10\%$.

For the three systems with highly flared observations, HSE masses could not be directly calculated. In another five cases, the morphology was found to be highly disturbed, such that the assumptions of HSE and spherical symmetry do not hold. In these cases, we extracted a single global temperature within a circular aperture of 300 kpc radius, and used the mass-temperature relation of \citet{Umetsu:2020} to estimate $R_{500}$. We then integrated the luminosity and the temperature within the corresponding apertures. The temperatures, luminosities, and HSE masses for the 44 groups that strictly match the X-GAP selection criteria are provided in Table \ref{tab:lxt}. For the observations that were strongly affected by soft proton flares, the masses are estimated from the $M_{500}-T_{300kpc}$ relation of \citet{Umetsu:2020}. The same applies to the merging systems, where the assumption of a single halo in hydrostatic equilibrium does not hold.

\section{Comparison with literature measurements}

\begin{figure*}
    \resizebox{\hsize}{!}{\includegraphics[width=0.5\textwidth]{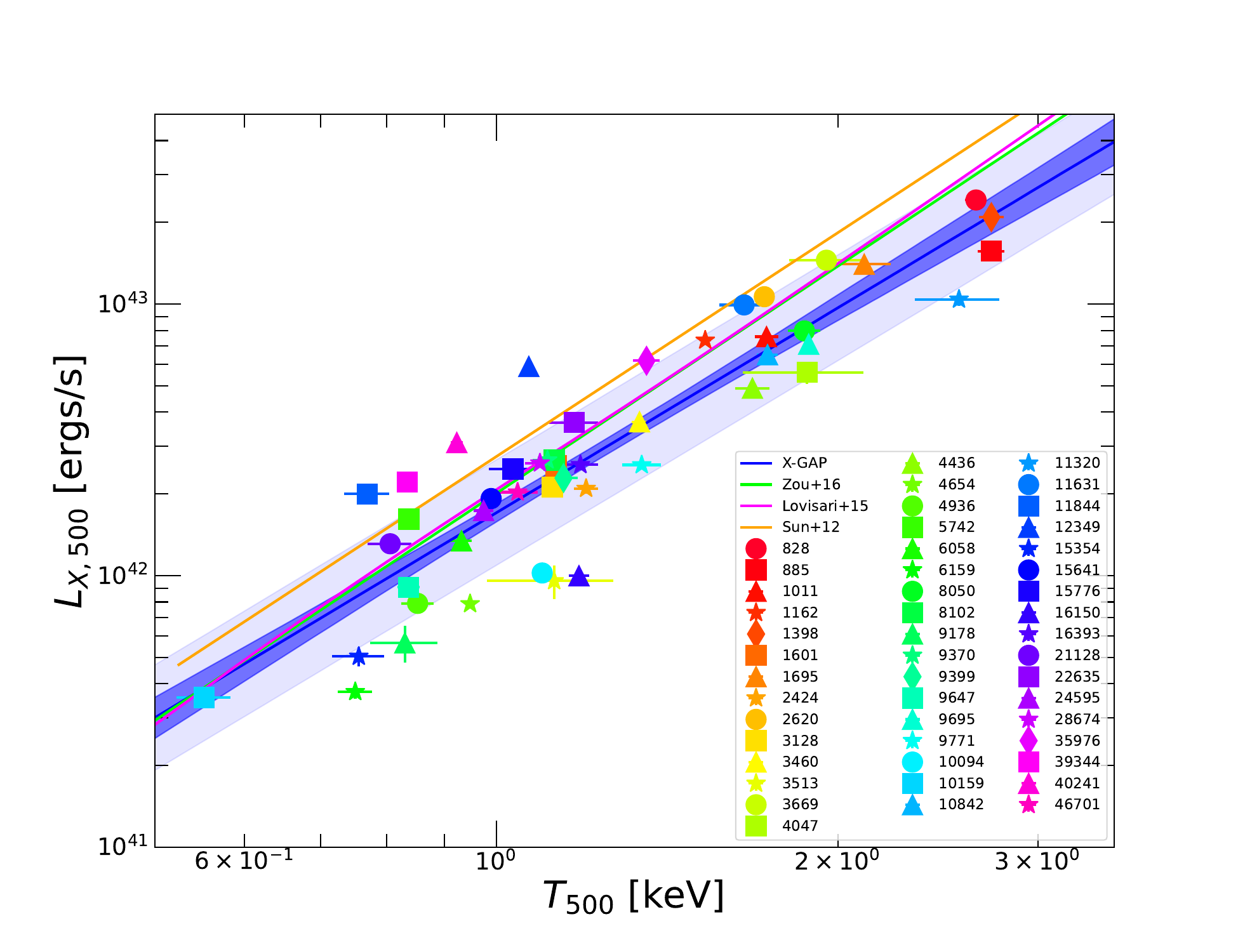}
    \includegraphics[width=0.5\textwidth]{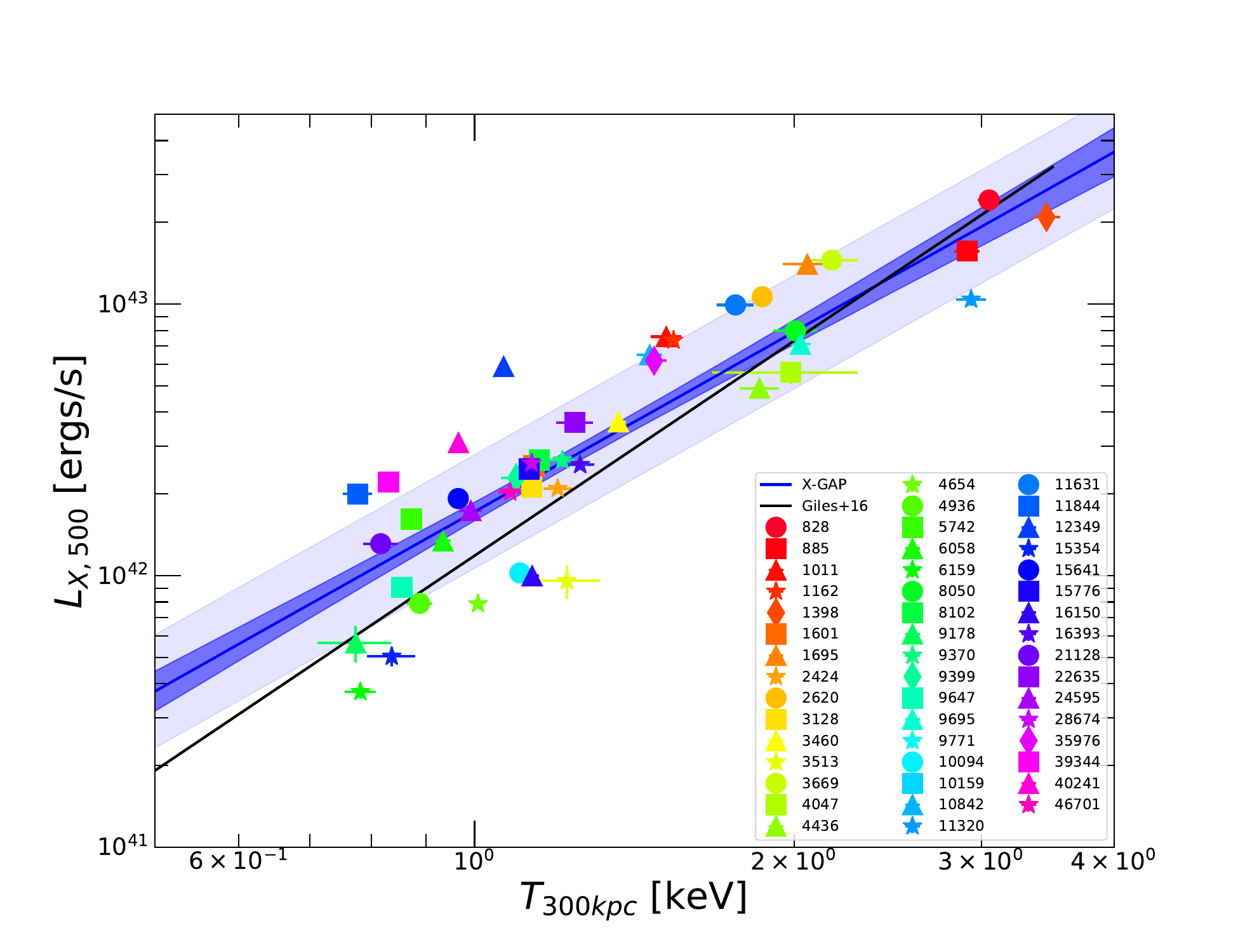}}
    \caption{\lxt~relation as a function of $T_{500}$ (left) and $T_{300kpc}$ (right) for X-GAP (coloured symbols, see Table \ref{tab:lxt}) in comparison with literature measurements. In the left-hand panel, the magenta and green curves show the bias-corrected relations from \citet{Lovisari:2015} and \citet{Zou:2016}, respectively. The orange curve in the right-hand panel is the $L_{X,500}-T_{300kpc}$ relation from XMM-XXL \citep{Giles:2016}. In both panels, the blue curve shows the fit to the X-GAP \lxt~relation, with the uncertainty in the mean indicated as the dark blue shaded area and the scatter around the mean indicated in light blue.}
    \label{fig:lxt_literature}
\end{figure*}

\begin{table}
\caption{Results of power-law fits to the \lxt~relation with Eq. \ref{eq:lxtfit}.}
    \label{tab:lxtfit}
    \centering
    \begin{tabular}{cccc}
    \hline
    Relation & A & B & $\sigma_{L|T}$\\
    \hline
    \hline
    $L_{0.5-2,500}-T_{500}$ & $0.24\pm0.04$ & $2.54\pm0.19$ & $0.20\pm0.02$\\
    $L_{0.5-2,500}-T_{300kpc}$ & $0.25\pm0.04$ & $2.18\pm0.18$ & $0.21\pm0.02$\\
    \end{tabular}
    
\end{table}

To check how the X-GAP \lxt~relation compares with previous estimates, we retrieved \lxt~estimates from the literature and plotted them against our measurements. Specifically, we considered the \lxt~relations at $R_{500}$ for ROSAT-selected groups, both for the brightest systems in the ROSAT all-sky survey \citep{Lovisari:2015} and the 400d area \citep{Zou:2016}. We also considered the \citet{Sun:2012} relation, which was extracted from a set of 43 groups in the \emph{Chandra} archive. The \citet{Lovisari:2015} relation in the [0.1-2.4] keV band was converted to the [0.5-2] keV  band assuming a single-temperature APEC model and a metallicity of $0.3Z_\odot$. We also compared with the \lxt~relation of groups selected in the XMM-XXL survey \citep{Giles:2016}. In the latter case, the temperatures are estimated within a fixed aperture of 300 kpc radius. The comparison with these relations can be found in Fig. \ref{fig:lxt_literature}. We also fitted the \lxt~relation of the X-GAP sample, without considering selection effects (see Sect. \ref{sec:selfunc}). To this aim, we modeled the relation as a power law with log-normal intrinsic scatter,
\begin{equation}
    \log\left(\frac{L_{0.5-2,500}}{10^{42}\mbox{ erg/s}}\right) = A + B \log\left(\frac{T}{\mbox{1 keV}}\right) \pm \sigma_{L|T} .\label{eq:lxtfit}
\end{equation}
We used \texttt{PyMC}\footnote{\href{https://www.pymc.io/welcome.html}https://www.pymc.io} to fit the relation to the data, including the uncertainties on both axes. The best fitting parameters for the $L_{0.5-2,500}-T_{500}$ and $L_{0.5-2,500}-T_{300kpc}$ relations are provided in Table \ref{tab:lxtfit} and shown in blue in Fig. \ref{fig:lxt_literature}.

We can see that the X-GAP \lxt~relation lies close to previous relations obtained from X-ray selected groups. Our relation agrees well with the \citet{Lovisari:2015} and \citet{Zou:2016} relations around 1 keV, but the slope retrieved here is slightly shallower. Conversely, the \lxt~relation from \citet{Sun:2012} has a higher normalisation, likely because of a biased group selection based on the \emph{Chandra} archive. The relatively flat slope is due to the most massive systems in the sample ($2-3$ keV), which appear under-luminous with respect to all the relations considered here. The flatter slope retrieved here can be largely explained by selection effects, as the most massive systems in the sample are expected to include primarily down-scattered systems, since the upper luminosity cut (Eq. \ref{eq:lxsel}) excludes brighter systems of similar mass from the sample. Our simulations (Sect. \ref{sec:selfunc}) show that this is the primary reason for the lower X-ray luminosity of the selected systems at the high-temperature end. Conversely, the X-GAP \lxt~relation appears close to the XMM-XXL relation \citep{Giles:2016} at the high-mass end and slightly above around 1 keV. The intrinsic scatter of the relation ($0.2$ dex) is similar to the value retrieved by \citet{Giles:2016} in XMM-XXL but nearly twice as high as the values estimated from the brightest ROSAT groups \citep{Lovisari:2015}. While we differ a full analysis of the X-GAP scaling relations including the selection function to a further paper, we note that all the relations lie within 0.15 dex of the X-GAP relation at all temperatures, which reinforces the tension with the predictions of the FLAMINGO \fgasm8~run.

\section{Data table}
\begin{table*}
\caption{\label{tab:lxt}Luminosities and temperatures of X-GAP groups considered in this study.}
\begin{center}
\begin{tabular}{ccccccc}
\hline
Group ID & $z$ & $T_{500}$ & $T_{300kpc}$ & $L_{X,0.5-2}$ & $M_{500,HSE}$ & $R_{500,HSE}$\\ 
 &  & keV & keV & $10^{42}$ erg/s & $10^{13}M_\odot$ & kpc \\ 
\hline
\hline
828 & 0.046 & $2.60_{-0.06}^{+0.06}$ & $3.05_{-0.07}^{+0.08}$ & $24.14_{-0.38}^{+0.37}$ & $12.88_{-1.26}^{+1.50}$ & $773_{-26}^{+29}$ \\ 
885 & 0.047 & $2.69_{-0.08}^{+0.08}$ & $2.91_{-0.08}^{+0.08}$ & $15.67_{-0.32}^{+0.29}$ & $16.25_{-2.61}^{+3.27}$ & $835_{-47}^{+53}$ \\ 
1011 & 0.046 & $1.69_{-0.04}^{+0.03}$ & $1.52_{-0.06}^{+0.05}$ & $7.58_{-0.13}^{+0.15}$ & $6.34_{-0.38}^{+0.46}$ & $610_{-12}^{+15}$ \\ 
1162 & 0.044 & $1.51_{-0.02}^{+0.02}$ & $1.54_{-0.02}^{+0.02}$ & $7.36_{-0.13}^{+0.15}$ & $10.11_{-1.22}^{+1.32}$ & $714_{-30}^{+30}$ \\ 
1398 & 0.046 & $2.67_{-0.07}^{+0.07}$ & $3.45_{-0.10}^{+0.10}$ & $20.90_{-0.42}^{+0.41}$ & $12.94_{-1.56}^{+1.81}$ & $774_{-32}^{+35}$ \\ 
1601 & 0.034 & $1.12_{-0.01}^{+0.01}$ & $1.14_{-0.01}^{+0.01}$ & $2.54_{-0.05}^{+0.04}$ & $3.97_{-0.22}^{+0.25}$ & $524_{-9}^{+11}$ \\ 
1695 & 0.039 & $2.08_{-0.12}^{+0.11}$ & $2.06_{-0.11}^{+0.11}$ & $14.02_{-0.60}^{+0.57}$ & $9.78_{-2.05}^{+3.14}$ & $707_{-53}^{+69}$ \\ 
2424 & 0.040 & $1.19_{-0.02}^{+0.03}$ & $1.20_{-0.04}^{+0.04}$ & $2.10_{-0.09}^{+0.09}$ & $2.54_{-0.23}^{+0.27}$ & $451_{-14}^{+16}$ \\ 
2620 & 0.039 & $1.69_{-0.02}^{+0.02}$ & $1.87_{-0.03}^{+0.03}$ & $10.65_{-0.13}^{+0.11}$ & $6.94_{-0.48}^{+0.55}$ & $631_{-15}^{+16}$ \\ 
3128 & 0.033 & $1.11_{-0.01}^{+0.02}$ & $1.13_{-0.02}^{+0.01}$ & $2.12_{-0.05}^{+0.05}$ & $2.32_{-0.18}^{+0.21}$ & $439_{-12}^{+13}$ \\ 
3460 & 0.043 & $1.33_{-0.02}^{+0.02}$ & $1.37_{-0.02}^{+0.03}$ & $3.69_{-0.09}^{+0.09}$ & $4.50_{-0.46}^{+0.68}$ & $545_{-19}^{+26}$ \\ 
3513$^\dagger$ & 0.036 & $1.11_{-0.12}^{+0.19}$ & $1.22_{-0.08}^{+0.11}$ & $0.97_{-0.14}^{+0.13}$ & $4.18_{-0.75}^{+0.84}$ & $533_{-32}^{+36}$ \\ 
3669$^\dagger$ & 0.048 & $1.95_{-0.15}^{+0.19}$ & $2.17_{-0.11}^{+0.15}$ & $14.48_{-0.63}^{+0.56}$ & $9.84_{-1.68}^{+1.80}$ & $706_{-40}^{+43}$ \\ 
4436 & 0.046 & $1.65_{-0.06}^{+0.06}$ & $1.85_{-0.08}^{+0.08}$ & $4.90_{-0.13}^{+0.14}$ & $6.10_{-1.02}^{+1.83}$ & $603_{-36}^{+55}$ \\ 
4654 & 0.022 & $0.94_{-0.01}^{+0.01}$ & $1.01_{-0.01}^{+0.01}$ & $0.79_{-0.02}^{+0.02}$ & $3.13_{-0.35}^{+0.42}$ & $486_{-19}^{+21}$ \\ 
4936 & 0.042 & $0.85_{-0.03}^{+0.03}$ & $0.89_{-0.03}^{+0.02}$ & $0.79_{-0.06}^{+0.07}$ & $1.85_{-0.29}^{+0.36}$ & $405_{-22}^{+25}$ \\ 
5742 & 0.034 & $0.84_{-0.02}^{+0.02}$ & $0.87_{-0.01}^{+0.01}$ & $1.62_{-0.04}^{+0.04}$ & $4.20_{-1.10}^{+1.54}$ & $534_{-51}^{+59}$ \\ 
6058 & 0.045 & $0.93_{-0.02}^{+0.02}$ & $0.93_{-0.02}^{+0.02}$ & $1.34_{-0.05}^{+0.05}$ & $2.39_{-0.46}^{+0.56}$ & $441_{-30}^{+32}$ \\ 
6159 & 0.024 & $0.75_{-0.03}^{+0.03}$ & $0.78_{-0.02}^{+0.03}$ & $0.37_{-0.02}^{+0.02}$ & $1.85_{-0.42}^{+0.65}$ & $408_{-34}^{+42}$ \\ 
8050 & 0.047 & $1.82_{-0.06}^{+0.05}$ & $2.01_{-0.09}^{+0.10}$ & $7.95_{-0.21}^{+0.21}$ & $7.67_{-0.82}^{+1.16}$ & $650_{-24}^{+32}$ \\ 
8102 & 0.033 & $1.11_{-0.02}^{+0.02}$ & $1.15_{-0.02}^{+0.02}$ & $2.66_{-0.07}^{+0.07}$ & $2.39_{-0.26}^{+0.36}$ & $443_{-17}^{+21}$ \\ 
9178 & 0.040 & $0.83_{-0.06}^{+0.06}$ & $0.77_{-0.10}^{+0.02}$ & $0.57_{-0.09}^{+0.08}$ & $2.19_{-0.65}^{+1.34}$ & $429_{-47}^{+75}$ \\ 
9370 & 0.038 & $1.09_{-0.03}^{+0.03}$ & $1.21_{-0.03}^{+0.04}$ & $2.67_{-0.13}^{+0.14}$ & $4.49_{-0.98}^{+1.69}$ & $545_{-43}^{+62}$ \\ 
9399 & 0.035 & $1.13_{-0.03}^{+0.03}$ & $1.09_{-0.03}^{+0.04}$ & $2.29_{-0.09}^{+0.09}$ & $3.14_{-0.50}^{+0.76}$ & $485_{-27}^{+36}$ \\ 
9647 & 0.023 & $0.83_{-0.02}^{+0.02}$ & $0.85_{-0.02}^{+0.02}$ & $0.91_{-0.03}^{+0.03}$ & $1.33_{-0.12}^{+0.16}$ & $365_{-11}^{+14}$ \\ 
9695 & 0.038 & $1.83_{-0.02}^{+0.02}$ & $2.03_{-0.04}^{+0.05}$ & $7.12_{-0.05}^{+0.05}$ & $6.96_{-0.19}^{+0.20}$ & $631_{-5}^{+6}$ \\ 
9771 & 0.044 & $1.35_{-0.06}^{+0.06}$ & $1.26_{-0.03}^{+0.03}$ & $2.56_{-0.10}^{+0.08}$ & $6.86_{-1.05}^{+1.09}$ & $627_{-34}^{+32}$ \\ 
10094 & 0.031 & $1.08_{-0.02}^{+0.02}$ & $1.10_{-0.02}^{+0.02}$ & $1.02_{-0.03}^{+0.04}$ & $2.38_{-0.26}^{+0.32}$ & $442_{-16}^{+20}$ \\ 
10159$^\dagger$ & 0.031 & $0.55_{-0.03}^{+0.04}$ & $0.36_{-0.04}^{+0.08}$ & $0.35_{-0.03}^{+0.02}$ & $0.67_{-0.15}^{+0.23}$ & $290_{-22}^{+34}$ \\ 
10842 & 0.040 & $1.69_{-0.04}^{+0.03}$ & $1.46_{-0.03}^{+0.05}$ & $6.50_{-0.10}^{+0.11}$ & $7.57_{-0.95}^{+1.38}$ & $649_{-28}^{+37}$ \\ 
11320$^\dagger$ & 0.045 & $2.51_{-0.15}^{+0.45}$ & $2.93_{-0.09}^{+0.09}$ & $10.41_{-0.51}^{+0.55}$ & $15.49_{-2.48}^{+2.47}$ & $823_{-44}^{+44}$ \\ 
11631$^\dagger$ & 0.046 & $1.63_{-0.08}^{+0.09}$ & $1.76_{-0.07}^{+0.07}$ & $9.93_{-0.35}^{+0.30}$ & $7.20_{-1.18}^{+1.18}$ & $637_{-35}^{+35}$ \\ 
11844 & 0.038 & $0.76_{-0.03}^{+0.04}$ & $0.78_{-0.03}^{+0.02}$ & $2.00_{-0.12}^{+0.11}$ & $2.52_{-0.77}^{+1.48}$ & $450_{-51}^{+75}$ \\ 
12349 & 0.036 & $1.07_{-0.01}^{+0.01}$ & $1.06_{-0.00}^{+0.00}$ & $5.90_{-0.04}^{+0.05}$ & $2.53_{-0.10}^{+0.10}$ & $451_{-6}^{+6}$ \\ 
15641 & 0.027 & $0.97_{-0.01}^{+0.01}$ & $0.97_{-0.02}^{+0.02}$ & $1.92_{-0.02}^{+0.02}$ & $2.83_{-0.48}^{+0.90}$ & $469_{-27}^{+46}$ \\ 
15776 & 0.036 & $1.03_{-0.04}^{+0.06}$ & $1.13_{-0.03}^{+0.03}$ & $2.47_{-0.18}^{+0.15}$ & $2.36_{-0.60}^{+1.12}$ & $441_{-41}^{+61}$ \\ 
16150 & 0.032 & $1.18_{-0.03}^{+0.03}$ & $1.13_{-0.02}^{+0.02}$ & $1.00_{-0.05}^{+0.05}$ & $4.81_{-0.96}^{+1.23}$ & $559_{-40}^{+44}$ \\ 
16393$^\dagger$ & 0.046 & $1.17_{-0.04}^{+0.05}$ & $1.26_{-0.04}^{+0.04}$ & $2.54_{-0.12}^{+0.12}$ & $4.34_{-0.69}^{+0.69}$ & $538_{-29}^{+29}$ \\ 
22635$^\dagger$ & 0.034 & $1.16_{-0.06}^{+0.07}$ & $1.24_{-0.05}^{+0.05}$ & $3.68_{-0.18}^{+0.17}$ & $4.29_{-0.71}^{+0.70}$ & $538_{-30}^{+29}$ \\ 
28674$^\dagger$ & 0.037 & $1.08_{-0.03}^{+0.05}$ & $1.13_{-0.02}^{+0.02}$ & $2.60_{-0.08}^{+0.07}$ & $3.73_{-0.58}^{+0.58}$ & $513_{-26}^{+26}$ \\ 
35976 & 0.036 & $1.34_{-0.03}^{+0.04}$ & $1.48_{-0.04}^{+0.04}$ & $6.18_{-0.19}^{+0.20}$ & $5.07_{-0.85}^{+0.95}$ & $569_{-34}^{+33}$ \\ 
39344 & 0.028 & $0.83_{-0.01}^{+0.01}$ & $0.83_{-0.01}^{+0.01}$ & $2.21_{-0.04}^{+0.04}$ & $2.23_{-0.19}^{+0.26}$ & $434_{-13}^{+16}$ \\ 
40241 & 0.049 & $0.92_{-0.01}^{+0.01}$ & $0.97_{-0.01}^{+0.01}$ & $3.09_{-0.06}^{+0.06}$ & $3.22_{-0.56}^{+0.89}$ & $487_{-30}^{+41}$ \\ 
46701 & 0.042 & $1.04_{-0.05}^{+0.05}$ & $1.08_{-0.03}^{+0.03}$ & $2.03_{-0.09}^{+0.09}$ & $1.74_{-0.43}^{+0.74}$ & $397_{-35}^{+50}$ \\ 
\hline
\end{tabular}
\end{center}\textbf{Column description:} 1: Group identifier in the \citet{Tempel:2017} catalogue. The $\dagger$ sign indicates that the mass was estimated from the $T_{300kpc}-M_{500}$ relation of \citet{Umetsu:2020} instead of the direct HSE mass. 2: Mean group redshift. 3: Mean temperature within $R_{500}$. 4: Mean temperature within a fixed aperture of 300 kpc. 5: HSE mass within an overdensity of 500 critical. 6: Overdensity radius $R_{500}$. 
\end{table*}

\end{document}